\documentclass[11pt]{article}

\usepackage{natbib}
\usepackage{setspace}
\usepackage{lineno}
\usepackage{graphics}
\usepackage{graphicx}
\usepackage{amsmath}
\usepackage{amssymb}
\usepackage{subfig} 
\usepackage{mathrsfs}

\newcommand{\vect}[1]{\mathbf{#1}}

\newcommand{\bogs}[1]{\boldsymbol{#1}}

\newcommand{\sfrac}[2]{{\textstyle\frac{#1}{#2}}}

\def\div{\mathrm{div}}

\begin{document}

\title{\bf Divergence Measures as Diversity Indices}

\author{Karim T. Abou--Moustafa\\       
       {\small Dept. of Computing Science, University of Alberta}\\
       {\small Edmonton, Alberta T6G 2E8, Canada}\\
       {\small\texttt{aboumous@cs.alberta.ca}}
      }
      
\date{Septembre 6th 2013}
      
\maketitle

\begin{abstract}
Entropy measures of probability distributions are widely used measures in ecology, biology, genetics, and in other fields, to quantify 
species diversity of a community.
Unfortunately, entropy--based diversity indices, or diversity indices for short, suffer from three problems.
First, when computing the diversity for samples withdrawn from communities with different structures, diversity indices can easily yield 
non-comparable and hard to interpret results.
Second, diversity indices impose weighting schemes on the species distributions that unnecessarily emphasize low abundant rare species, or 
erroneously identified ones.
Third, diversity indices do not allow for comparing distributions against each other, which is necessary when a community has a well-known 
species' distribution.

In this paper we propose a new general methodology based on information theoretic principles to quantify the species diversity of a community.
Our methodology, comprised of two steps, naturally overcomes the previous mentioned problems, and yields comparable and easy to interpret
diversity values.
We show that our methodology retains all the functional properties of any diversity index, and yet is far more flexible than entropy--based 
diversity indices.  
Our methodology is easy to implement and is applicable to any community of interest.
\end{abstract}



\section{Introduction}
\label{sec:introduction}

Let $\mathcal{C}$ be a community of living organisms where each member of this community (called an individual) has the label of a species.
Let $s$ be the number of different species (or individual categories) in $\mathcal{C}$, where the species are labelled from $1$ to $s$.
Denote the probabilities of species discovery, or relative abundance, by $\bogs{\pi} = [\pi_1,\pi_2,\dots,\pi_s]^\top$, where
$\sum_{j=1}^s \pi_j = 1$, and $\pi_j \geq 0$.
Suppose a random sample of $m$ individuals is taken from $\mathcal{C}$ and each individual is correctly classified according to its species 
identity.
If $x_j$ is the number of individuals of the $j$th species observed in the sample, then $\vect{x} = [x_1,x_2,\dots,x_s]^\top$, 
where $\sum_{j=1}^s x_j = m$, is a multinomial distribution $\mathcal{M}$ with parameters $ (m,\bogs{\pi})$;
or $\vect{x} \sim \mathcal{M}(\bogs{\pi})$ for short.

The diversity of community $\mathcal{C}$ is a key concept in ecological studies. 
The main difficulty in measuring the (self) diversity of a community (or $\alpha$--diversity) is compressing the complexity of a distribution, 
with a multidimensional representation of species relative abundance, into a single scalar statistic \citep{kelvin_2012}.
In its simplest definition, a diversity index is a function of two properties that characterize the species in $\mathcal{C}$:
(\emph{i}) the number of species present in the community (species \emph{richness} or \emph{abundance}), and 
(\emph{ii}) the evenness with which the individuals are distributed among these species 
(species \emph{relative evenness} or \emph{equitability}).
If $s$ is the number of species in $\mathcal{C}$, then the diversity is higher whenever $s$ is increasing, \emph{and/or} 
$\mathcal{M}(\bogs{\pi})$ approaches the uniform distribution $\mathcal{U}$; 
i.e., $\pi_i \approx \pi_j$ for $1 \leq i,j \leq s$ and $i \neq j$. 

The previous verbal definition of diversity, although based on \emph{``ecological''} concepts, naturally coincides with the definition of 
entropy in information theory \citep{shannon_1948}.
Indeed, plant, animal, and microbial ecologists have heavily relied on entropy measures as diversity indices.
Further, each research community has proposed its own variants of diversity measures, each exhibiting different sensitivity to one of the 
aspects characterizing the community (richness, evenness, etc.).
Despite the plethora of these diversity indices, the ubiquitous Shannon (or Shannon--Wiener) entropy \citep{shannon_1948} 
seems to be the index of choice for various ecology researchers\footnote{Other diversity indices will be discussed in the following sections.}.
A widely used estimator for Shannon's entropy $H$ is the maximum likelihood estimate (MLE) given by:
\begin{eqnarray}
\widehat{H} & = & -\sum_{j=1}^s \widehat{\pi}_j \log_2 (\widehat{\pi}_j) = - \sum_{j=1}^s \sfrac{x_j}{m} \log_2 (\sfrac{x_j}{m}), \label{eq:shannon_entropy}
\end{eqnarray}
where $\widehat{\pi}_j$ is the MLE of $\pi_j$.
Note that $H$, like any other entropy measure, is a function defined on the space of distribution functions satisfying some postulates:
(\emph{i}) non negativity, 
(\emph{ii}) attains a maximum for the uniform distribution, and 
(\emph{iii}) has a minimum when the distribution is degenerate.
Thus a measure of entropy is in fact, an index of similarity of a distribution function with the uniform distribution $\mathcal{U}$.

\begin{figure}[t]
\centering
\includegraphics[scale=0.5]{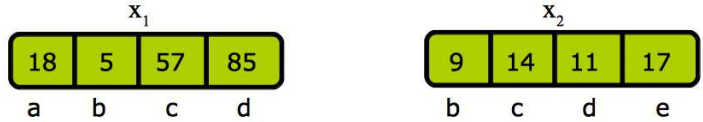} 
\caption{In this example, and using Equation (\ref{eq:shannon_entropy}), the entropies for samples $\vect{x}_1$ and $\vect{x}_2$ are: 
$1.52$ and $1.96$, respectively. Although it is possible to conclude that $\vect{x}_2$ is more diverse than $\vect{x}_1$, one should note that
these two samples are not comparable since the common species between both samples are only `b', `c', and `d' but not `a' nor `e'.}
\label{fig:ex1}
\end{figure}

In this paper, we consider three general problems of entropy--based diversity measures, exemplified by Shannon's entropy, when used to 
compare the diversity between two or more communities.
Note that the following discussion is not restricted to any particular type of communities.

The first problem that affects the comparison of multiple communities is due to the convex weight $\widehat{\pi}_j$ assigned to the log term in 
Equation (\ref{eq:shannon_entropy}), thereby assigning a larger weight per individual to rare than common species.
Such a weighting scheme will increase the influence of rare species while decrease the influence of common species, thereby creating a 
balance between rare and common species.
While such a weighting scheme might be useful in some cases, we argue whether it is always desirable.
For instance, if some of the rare species are not the usual habitants of a community, i.e., noisy samples, or some individuals were not 
correctly classified to their true species identity, then $H$ will unnecessarily emphasize the importance of such samples.
More importantly, the reader should note that this weighting scheme alters the true distribution of the species. 
Thus, it would be desirable to have the flexibility of computing the diversity of $\mathcal{C}$ without relying on such weights.

The second problem arises when comparing two or more values of the Shannon index.
That is, when comparing the diversity of two samples, and each collected from a different community, if the two samples do not contain the 
same species categories and all their relative abundances are non--zeros, Shannon's entropy will be a misleading index of the diversity of 
both communities.
The reason for that is that Shannon's entropy positively correlates with species richness (the number of species categories) and evenness.
To see this, consider the example depicted in Figure (\ref{fig:ex1}).
In this example, $\vect{x}_1$ and $\vect{x}_2$ are two samples withdrawn from communities $\mathcal{C}_1$ and $\mathcal{C}_2$, respectively.
Using Equation (\ref{eq:shannon_entropy}), the entropies for $\vect{x}_1$ and $\vect{x}_2$ are $1.52$ and $1.96$, respectively.
Although, at first glance, it is possible to conclude that $\mathcal{C}_2$ is more diverse than $\mathcal{C}_1$, one should note that these 
two values are not comparable since the common species between both samples are only `b', `c', and `d' but not `a' nor `e'.
In fact, it is enough to have one different species in both samples to render the values not comparable.
Note that the value of $H$ in the examples above will be more perplexing if the number of species in both samples are not equal, and the 
situation becomes worse when there are tens or hundreds of samples to compare, each with hundreds or thousands of species.

\begin{figure}[t]
\centering
\subfloat[Step 1]{\includegraphics[scale=0.35]{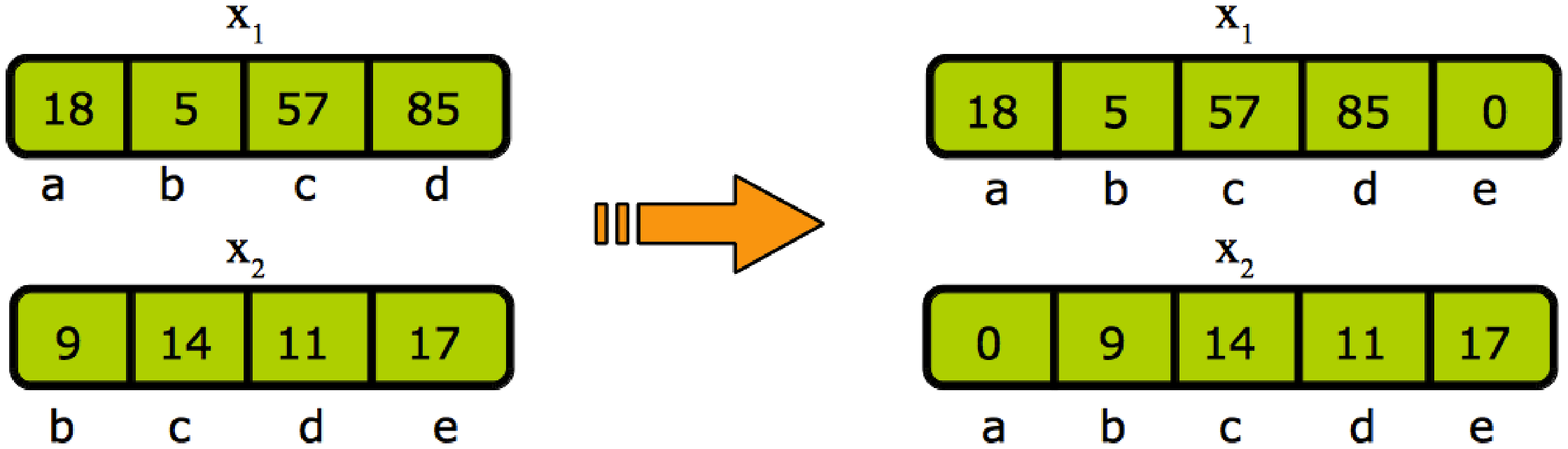} 
\label{fig:ex4}
}\quad\quad
\subfloat[Step 2]{\includegraphics[scale=0.35]{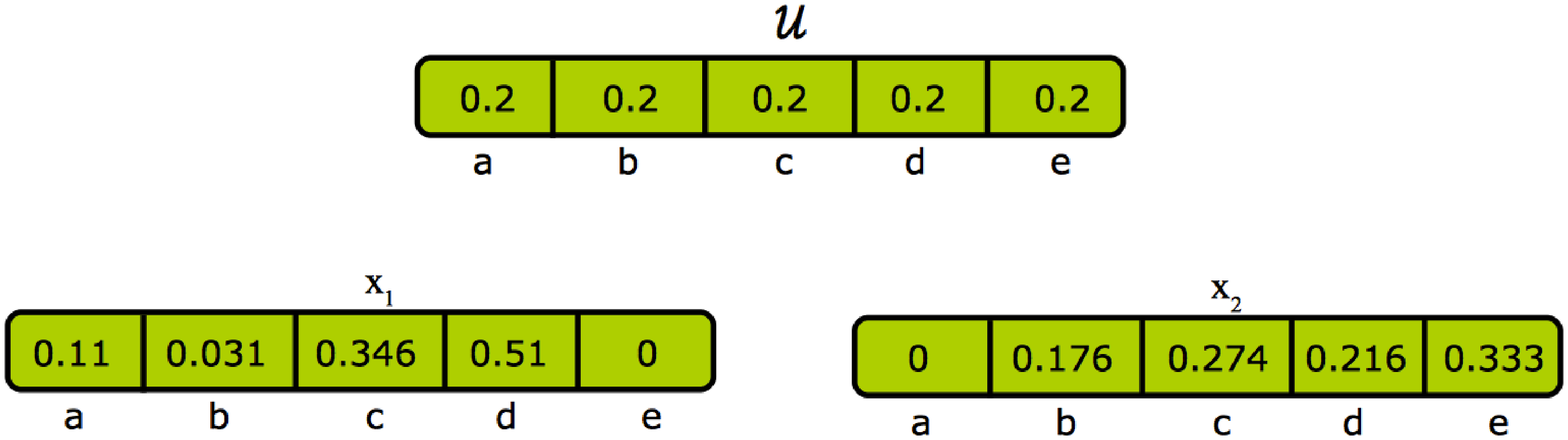} 
\label{fig:ex5}} 
\caption{{\bf (a)} The first step of our proposed approach creates a set of species that is the union of all species from $\vect{x}_1$ 
and $\vect{x}_2$. Then, $\vect{x}_1$ and $\vect{x}_2$ are re-presented using the unified set of species. Note that this will introduce zero 
counts in the new representation. Since $0 \log 0 = 0$, using Shannon's entropy for the new $\vect{x}_1$ and $\vect{x}_2$ will be identical 
to the situation in Figure (\ref{fig:ex1}). However, using the divergence naturally overcomes this problem.
{\bf (b)} Samples $\vect{x}_1$ and $\vect{x}_2$ are represented using their distributions (or relative normalized abundances), and 
$\mathcal{U}$ is the uniform distribution over the unified set of species.
Here we have used $\mathcal{U}$ as the reference distribution to illustrate the main idea.
In the second step, the proposed framework measures the diversity of $\vect{x}_1$ and $\vect{x}_2$ as the divergence between 
$\vect{x}_1$ and $\mathcal{U}$, and between $\vect{x}_2$ and $\mathcal{U}$, respectively.
} 
\label{fig:ex4ex5}
\end{figure}

The third problem is due to the definition of entropy itself which turns to limit the scope of diversity.
First, based on the definition of entropy, note that computing the diversity of $\mathcal{C}$ is equivalent to measuring the similarity 
between the distribution of species in $\mathcal{C}$ and the uniform distribution $\mathcal{U}$ over the same set of species.
Second, note also that $\mathcal{U}$ has the highest entropy (or diversity) among all other possible distributions defined over the $s$ 
species of $\mathcal{C}$.
These two remarks imply that $\mathcal{U}$ is the ultimate reference distribution for comparisons for any community $\mathcal{C}$.
However, in nature surrounding us, it is less probable that a community of any living species can have such a uniform distribution.
It is more reasonable to believe that each community will have a latent distribution $\mathcal{M}(\bogs{\pi}^*)$ that is not necessarily 
uniform.
Biologists, after a fair amount of research, may provide a reasonable estimate or a model $\mathcal{M}(\bogs{\widehat{\pi}}^*)$ for the 
latent distribution, which makes it the new reference distribution for a given type of communities instead of $\mathcal{U}$.
For instance, in macroecology and community ecology, this is know as the occupancy frequency distribution (OFD) and there has been many
advances in that regard since it was first introduced by Raunki{\ae}r in 1918 \citep{gaston_2002,hui_2010}.  
In such cases, using $\mathcal{M}(\bogs{\widehat{\pi}}^*)$ as a reference distribution will be preferable over using $\mathcal{U}$.
Further, if $\mathcal{M}(\bogs{\widehat{\pi}}_1)$ and $\mathcal{M}(\bogs{\widehat{\pi}}_2)$ are empirically estimated from two other 
communities $\mathcal{C}_1$ and $\mathcal{C}_2$, respectively, an interesting question then is, how to measure the pairwise 
similarity/dissimilarity directly between $\mathcal{M}(\bogs{\widehat{\pi}}_1)$, $\mathcal{M}(\bogs{\widehat{\pi}}_2)$, 
and $\mathcal{M}(\bogs{\widehat{\pi}}^*)$ without relying on their entropies? 

To overcome the aforementioned problems, we propose a new methodology for assessing and comparing the diversity of multiple communities.
Our approach, which is also grounded on information theoretic principles, has two steps.
In the first step, depicted in Figure (\ref{fig:ex4}), we overcome the problem of communities with different species by first defining a new 
set of species that is the union of all species from all communities under consideration.
Next, each community is re-represented using the new unified set of species, thereby creating a common ground for comparisons for all 
communities under study.
Note that, for Figure (\ref{fig:ex5}), using the new representation will introduce species with zero counts in the sample. 
If entropy is used to assess the diversity of these communities, then zero count species will be neglected by $H$ since $0 \log 0$ is 0,
which reduces to the problem depicted in Figure (\ref{fig:ex1}).
We overcome this problem, however, using the second step of our proposed framework. 

In the second step, we generalize entropy--based diversity indices to divergence--based indices.
That is, instead of measuring the entropy for each community given its new representation based on the unified set of species, we compute the 
divergence between the distribution of each community and the reference model $\mathcal{M}(\widehat{\bogs{\pi}}^*)$.
When $\mathcal{M}(\widehat{\bogs{\pi}}^*)$ is not known for the community under consideration, then one has no other option but to use the 
ultimate diverse distribution which is the uniform distribution $\mathcal{U}$ defined over the unified set of species, as depicted in Figure 
(\ref{fig:ex5}).
Unlike entropy--based measures, the divergence measures the dissimilarity (or difference) between any two probability distribution functions 
defined over the same set of outcomes.
In other words, the divergence between two distribution functions is analogous to the distance between two points in an Euclidean space.
As will be explained in \S~\ref{sec:divergence_based_diversity}, zero count species are not neglected by divergence measures, and they 
increase the dissimilarity between the two distributions.
Hence, by definition, the divergence overcomes the second and third problems of entropy--based measures mentioned above.
Further, divergence measures do not impose any weights that alter the original sample distribution under consideration, and therefore they 
also overcome the first problem we discussed above of entropy--based measures.

Readers familiar with Whittaker's {\em beta diversity} \citep{whittaker_1960} should note the difference between this type of diversity on 
one hand, and the methodology proposed here on the other hand. 
Beta diversity \citep[p. 320]{whittaker_1960} measures the extent of change in community composition, or degree of community differentiation, 
in relation to a complex-gradient of environment, or a pattern of environments.
Note that this description covers two different aspects for a community:
(\emph{i}) the change in the composition of the community itself, and
(\emph{ii}) the degrees of differences in diversity between the community itself (as a subgroup), its surrounding communities 
(other subgroups), and the species diversity at the regional or landscape scale.
See \citep{tuomisto_2010} for a clear overview of beta diversity.
Our proposed methodology as described above, is not addressing the extent of compositional change in one community, nor is addressing the 
relation and structural differences between a community and its surrounding communities, or its surrounding region at large.
To the best of our knowledge, we are unaware of any research in the literature that has addressed the above issues together with a proposed
solution.


\section{Overview of Diversity Indices}
\label{sec:diversity_indices}

Since its introduction in 1943 \citep{fisher_williams_1943,mcarthur_1955,margalef_1958}, the concept of species diversity has been defined in 
various and disparate ways leading to a plethora of diversity measures with different and rather ``conflicting'' characteristics 
\citep{jost_2006}.
This has led some researchers in the 70's, such as \citet{hurlbert_1971}, to conclude that species diversity is meaningless.
More recently, this debate has evolved to the need for a consistent terminology for quantifying species diversity 
\citep{moreno_2010,hanna_2010}.
The first effort to disambiguate the term is due to \citet{whittaker_1972}, followed by \citet{hill_1973}, and more recently by 
\citet{jost_2006}.
Most researchers, including Hurlbert, have agreed that the definition of a community's diversity within itself ($\alpha$--diversity) 
should, at best, be restricted to the one introduced in \S~\ref{sec:introduction}.
\citet{jost_2006} made a further distinction between a diversity index, such as $H$, and a diversity number.
In his argument: ``\emph{A diversity index is not necessarily a diversity. 
The radius of a sphere is an index of its volume but is not itself the volume, and using the radius in place of volume in engineering 
equations will give catastrophic misleading results}''.
Based on his argument, the diversity of a community reduces to finding a community that is composed of equally common species.
Using simple algebra, he devises an algorithm for recovering the diversity number given the value of a diversity index.
For instance, the expression for the diversity number based on Shannon's index is $\exp(-H)$.

In the literature, there are two other well known indices, the Simpson's index \citep{simpson_1949}:
$\mathit{Sp} = \sum_{j = 1}^s \widehat{\pi}_j^2$,
and the Chao-1 index \citep{chao_1984}:
$\mathit{Ch} = s + \frac{a^2}{2b}$,
where $a$ is the number of singletons (species with a single occurrence),
and $b$ is the number of doubletons (species with a double occurrences) in $\mathcal{C}$.
Simpson's index is sensitive to the abundance of the more plentiful species in a sample and therefore can be regarded as a measure 
of dominance concentration.
Similar to $H$, Simpson's index is a weighted mean of the relative abundances, and both measures were shown to be special cases from 
R\'enyi's entropy.
\cite{hill_1973} and \citet{jost_2006}, however, advised to use the reciprocal of Simpson's index, $1/\mathit{Sp}$, or the 
generalized entropy, $\ln(\mathit{Sp})$, as diversity numbers, while \cite{whittaker_1972} and \citep{pielou_1967} favoured 
the Gini--Simpson index: $1 - \mathit{Sp}$.

Shannon's and Simpson's indices perform as expected when approximating the diversity of common species, however each may fall short as a 
single complete measure when examining numerous low abundant organisms that dominate the composition of a community \citep{kelvin_2012}.
Both indices have been shown by Hill, through Renyi's definition of generalized entropy \citep{renyi_1960}, to have similar characteristics, 
but differing only in the contribution of low abundant species to the magnitude of the calculated statistic.
Renyi's entropy unifies Shannon's and Simpson's diversity indices as entropies with a parmeter $q$, the power to which the contribution
of taxonomic abundances is raised:
\begin{eqnarray}
D_q & = & \left( \sum_{i=1}^s \pi_i^q \right)^{\sfrac{1}{1-q}}. \label{eq:renyi}
\end{eqnarray}
Hence, $q$ values of 2, 1, and 0, are associated with Simpson's index, Shannon's index, and the total number of species detected, 
respectively.
While these are known as Hill numbers, surprisingly, Jost's interpretation and algorithm for recovering the diversity number from any 
entropy--based diversity index yields exactly the expression in Equation (\ref{eq:renyi}).

Chao-1 index, in fact, is a richness estimator -- i.e., an estimator for $s$ -- although various studies have used it as a diversity measure.
Chao-1 relies on the existence of singletons and doubletons in the sample. 
If no singletons nor doubletons in the sample, Chao-1 equals the number of observed species in the sample.
Note that Chao-1 does not strictly follow our chosen definition of diversity introduced in \S~\ref{sec:introduction} since it does not 
address the equitability of relative abundances in the sample.

Despite the differences between all the above indices, it is worth noting that various researchers consider that the number of species, 
Simpson's index, and Shannon's index are in some sense, similar evaluations for the number of species present in the sample, and they only 
differ in their propensity to include or exclude the relatively rare species \citep{hill_1973}.

In a different research path, Chao and Shen \citep{chao_shen_2003} consider three shortcomings of the MLE for $H$ in Equation 
(\ref{eq:shannon_entropy}):
(\emph{i}) Equation (\ref{eq:shannon_entropy}) is derived under the assumptions that $s$ is known,
(\emph{ii}) it is assumed that $m > s$, and
(\emph{iii}) the fact that the MLE $\widehat{\pi}_j$ is negatively biased; i.e., $\widehat{H}$ is an underestimate for $H$.
In practice, the true value of $s$ is unknown, and rare species may not be discovered in a sample due the existence of numerous 
low abundant species.
Further, due to negative bias of $\widehat{\pi}_j$, $\widehat{H}$ yields an estimation error that will differ between samples, depending on the 
diversity and evenness in each, and will be large for small samples \citep{hill_moffett_2003}.
Hence the authors proposed a nonparametric estimator for $H$ for the particular case when $s$ is unknown, while taking into account the 
possibility of having unseen species.
Note that the motivations for the Chao and Chen estimator are different from our motivations discussed in \S~\ref{sec:introduction}.
Further, their estimator relies on the concept of sample coverage to adjust the sample fraction for unseen species which relies on the 
presence of singletons and doubletons as in the Chao-1 index.
Such assumptions on singletons and doubletons might not be applicable in some domains.


\section{Divergence--based Diversity Measures}
\label{sec:divergence_based_diversity}

In this section we introduce our two-step framework for measuring the diversity using divergence measures.
We begin our discussion with the necessary notations.
Let $\{\mathcal{C}_i\}_{i=1}^n$ be the set of communities under study, and $\vect{x}_i = [x_i^1,\dots,x^j_i,\dots,x_i^{s_i}]^\top$ be the 
sample withdrawn from $\mathcal{C}_i$, where $s_i$ is the number of observed species (or OTUs) in $\mathcal{C}_i$.
Accordingly, $\vect{x}_i \sim \mathcal{M}(m_i,\bogs{\pi}_i)$, where $\sum_{j=1}^{s_i} x^j_i = m_i$ is the total number of individuals in the 
sample $\vect{x}_i$.
Let $\Omega_i = \{o_1,\dots,o_j,\dots,o_{s_i}\}$ be the set of species' labels (or OTUs) found in $\mathcal{C}_i$.
To avoid any reliance on the order of species labels in $\Omega_i$, for any label $o$, we use the following notation to index the elements 
of sample $\vect{x}_i$:
\begin{eqnarray}
\vect{x}_i(o) & = & 
\left\{ 
\begin{array}{l l}
x^j_i & \quad \text{if $o = o_j$ and $o_j \in \Omega_i$}, \\
0     & \quad \text{otherwise.}
\end{array}
\right.\label{eq:indexing_counts}
\end{eqnarray}

The first step of our proposed framework is to have a unified representation for all samples. 
To achieve this, let $\Omega^*$ be the union set of species collected from all the samples under consideration:
\begin{eqnarray}
\Omega^* & = & \bigcup_{i=1}^n \Omega_i ~ \equiv ~ \{o_1,\dots,o_s\},
\end{eqnarray}
where the cardinality of $\Omega^*$ is $s$.
The set $\Omega^*$ includes all $\{\Omega_i\}_{i=1}^n$, and hence all samples $\{\vect{x}_i\}_{i=1}^n$ need to be represented in terms of 
its elements.
This can be obtained using our notation for indexing the elements of $\vect{x}_i$ in Equation (\ref{eq:indexing_counts}):
\begin{eqnarray}
\bar{\vect{x}}_i & = & [\vect{x}_i(o_1),\vect{x}_i(o_2),\dots,\vect{x}_i(o_s)]^\top, ~ 1 \leq i \leq n,
\end{eqnarray}
where $\bar{\vect{x}}_i$ is the new sample representing $\mathcal{C}_i$ using $\Omega^*$.
Further, we define the empirical discrete distribution $\mathcal{X}_i$ from $\bar{\vect{x}}_i$ as:
\begin{eqnarray}
\mathcal{X}_i & = & \left[\widehat{\pi}_i^1,\dots,\widehat{\pi}_i^s\right]^\top ~ \equiv ~ 
					\left[\sfrac{\vect{x}_i(o_1)}{m_i},\cdots,\sfrac{\vect{x}_i(o_s)}{m_i}\right]^\top,
					~ 1 \leq i \leq n. \label{eq:all_xi_dist}
\end{eqnarray}

The rational for using $\Omega^*$ instead of $\{\Omega_i\}_{i=1}^n$ is that it provides a common ground for comparing all samples from 
different communities.
That is, it reduces the comparison between communities to the differences in the distribution of relatives abundances. 
The problem, however, is that the new representation $\bar{\vect{x}}_i$, and consequently the discrete distribution $\mathcal{X}_i$, 
is sparse; i.e., it contains a considerable number of zero elements since not all species in $\Omega^*$ are present in all $\mathcal{C}_i$'s.
Recall that entropy--based diversity measures correlate with the number of (nonzero) species in the sample, and with the evenness 
(or equitability) of the relative abundances (or the individuals' distribution in a sample).
When using entropy--based diversity measures on such representations, it is enough to have one zero element per sample (in any location) to 
render the entropy values meaningless and not comparable. 
This is exactly the scenario depicted in Figure (\ref{fig:ex4}), and since $0 \log 0 = 0$, it reduces to the problem in Figure 
(\ref{fig:ex1}).
Even if $\mathcal{X}_i$ does not have any zero elements, entropy--based measures will alter the original distribution to create a balance
between rare and abundant species.
In addition, entropy--based measures are not flexible in terms of the reference distribution, nor they allow for pairwise comparisons between
all samples.
We overcome these problem, however, using the second step of our proposed framework.


\subsection{From Entropy to Divergence}
\label{subsec:from_entropy_to_divergence}

To overcome the above problem, we rely on the basic definition of entropy (which coincides with our definition of diversity).
That is, an entropy measure is a function defined on the space of distribution functions satisfying some postulates:
(\emph{i}) non negativity, 
(\emph{ii}) attains a maximum for the uniform distribution (i.e., maximum diversity), and 
(\emph{iii}) has a minimum when the distribution is degenerate.
Thus a measure of entropy is in fact, an index of similarity of a distribution function with the uniform distribution $\mathcal{U}$.
Let us define the uniform discrete distribution over $\Omega^*$:
\begin{eqnarray}
\mathcal{U} & = & [u_1,u_2,\dots,u_s]^\top ~ = ~ [\sfrac{1}{s},\sfrac{1}{s},\cdots,\sfrac{1}{s}]^\top. \label{eq:uniform} 
\end{eqnarray}

The second step of our proposed framework is to replace the entropy of a distribution with a surrogate function that measures the 
dissimilarity between the given distribution, say $\mathcal{X}_i$ , and the reference distribution $\mathcal{M}(\bogs{\widehat{\pi}}^*)$.
When $\mathcal{M}(\bogs{\widehat{\pi}}^*)$ is not known, then one has no other option but to use the uniform distribution $\mathcal{U}$ defined 
over $\Omega^*$ as a reference distribution.

The natural function that measures the dissimilarity between any two probability distributions is the divergence, 
Ali--Silvey distance \citep{ali_silvey_66}, or $f$-divergence according to Csiszar \citep{csiszar_67,book_kullback_59}.
If $\mathfrak{D}$ is the space of probability distributions, and $\mathcal{P},\mathcal{Q} \in \mathfrak{D}$ are two distributions defined 
over the same set of outcomes $\mathfrak{E}$, then the divergence quantifies how $\mathcal{P}$ diverges from $\mathcal{Q}$ over all the 
elements of $\mathfrak{E}$.
For simplicity, the divergence between two probability distributions is analogous, for instance, to the Euclidean distance between two points 
in an Euclidean space.
The smaller the divergence between two distributions, the more similar these two distributions are, and vice versa.

The divergence between $\mathcal{P}$ and $\mathcal{Q}$, denoted by $\div(\mathcal{P},\mathcal{Q})$, has to satisfy some conditions.
One of the conditions relevant to our discussion is that $\div$ should be zero when $\mathcal{P} = \mathcal{Q}$, and as large as possible 
when $\mathcal{P}$ and $\mathcal{Q}$ are completely different.
The divergence by definition does not need to be symmetric, nor does it need to satisfy the triangle inequality, and hence it is different 
from distance metrics in that regard.
However, in this research work, we will consider symmetric divergence measures, and some will satisfy the triangle inequality.
That is, for $\mathcal{P}, \mathcal{Q}, \mathcal{Z} \in \mathfrak{D}$, all defined over $\mathfrak{E}$, then 
$\div(\mathcal{P},\mathcal{Q}) = \div(\mathcal{Q},\mathcal{P})$, and
$\div(\mathcal{P},\mathcal{Z}) \leq \div(\mathcal{P},\mathcal{Q}) + \div(\mathcal{Q},\mathcal{Z})$.

Since we are interested in discrete probability distributions, 
let $\mathcal{P} = [p_1,\dots,p_s]^\top$, and $\mathcal{Q} = [q_1,\dots,q_s]^\top$, where for $1 \leq j \leq s$,
$p_j \geq 0$, $q_j \geq 0$, $\sum_{j=1}^s p_j = 1$, and $\sum_{j=1}^s q_j = 1$.
For the purpose of measuring the diversity of a distribution, we shall consider the following divergence measures:
\begin{enumerate}
\item The total variational distance (or the $L_1$ distance) \citep{ali_silvey_66,csiszar_67}:
	\begin{eqnarray}
	D_V (\mathcal{P},\mathcal{Q}) & = &	\frac{1}{2} \sum_{j=1}^s |p_j - q_j|. \label{eq:div_l1}
	\end{eqnarray}

\item The Hellinger distance \citep{rao_95}:
	\begin{eqnarray}
	D_H (\mathcal{P},\mathcal{Q}) & = & \frac{1}{\sqrt{2}} \sum_{j=1}^s \left( \sqrt{p_j} - \sqrt{q_j} \right)^2. \label{eq:div_hell}
	\end{eqnarray}

\item The symmetric Kullback-Leibler (KL) divergence \citep{book_kullback_59}:
	\begin{eqnarray}
	D_{\mathrm{SKL}} (\mathcal{P},\mathcal{Q}) & = &	\sum_{j=1}^s \left( p_j - q_j \right) \log_2 \frac{p_j}{q_j}. \label{eq:div_skl}
	\end{eqnarray}

\item The Bhattacharyya distance \citep{bhattacharyya_43}:
	\begin{eqnarray}
	D_B (\mathcal{P},\mathcal{Q}) & = &	-\log \left( \sum_{j=1}^s \sqrt{p_j q_j} \right). \label{eq:div_bhatt}
	\end{eqnarray}

\item The square root of Jensen-Shannon divergence \citep{lin_jensenshannon_91}:
	\begin{eqnarray}
	D_{\mathrm{JS}} (\mathcal{P},\mathcal{Q}) & = &	\sqrt{\sfrac{1}{2} \div_\mathrm{KL}(\mathcal{P},\mathcal{Z}) + 
	                                                \sfrac{1}{2}\div_\mathrm{KL}(\mathcal{Q},\mathcal{Z})}, \label{eq:div_js} \\
	\div_\mathrm{KL}(\mathcal{P},\mathcal{Z}) & = & \sum_{j=1}^s p_j \log \frac{p_j}{z_j} , \nonumber \\
	\div_\mathrm{KL}(\mathcal{Q},\mathcal{Z}) & = & \sum_{j=1}^s q_j \log \frac{q_j}{z_j} , \nonumber 
	\end{eqnarray}
	where $\mathcal{Z} = \sfrac{1}{2}(\mathcal{P} + \mathcal{Q}) = \sfrac{1}{2}[p_1 + q_1,\dots,p_s + q_s]^\top$ 
	is the middle distribution for $\mathcal{P}$ and $\mathcal{Q}$, and
	$\div_\mathrm{KL}$ is the directed KL divergence \citep{book_kullback_59} between two distributions.
	All measures in Equations (\ref{eq:div_l1}) -- (\ref{eq:div_js}) have the following properties:
	(\emph{i}) $\div(\mathcal{P},\mathcal{Q})\geq 0$, 
	(\emph{ii}) $\div(\mathcal{P},\mathcal{P}) = 0$,
	(\emph{iii}) $\div(\mathcal{P},\mathcal{Q})=0$ iff $\mathcal{P}=\mathcal{Q}$, and
	(\emph{iv}) symmetry.
	Only $D_H$ and $D_\mathrm{JS}$ satisfy the triangle inequality.
	Note that both $D_H$ and $D_B$ are derived from the Bhattacharyya coefficient 
	$\Gamma(\mathcal{P},\mathcal{Q}) = \sum_{j=1}^s \sqrt{(p_jq_j)}$, where
	$D_H = 1 - \Gamma(\mathcal{P},\mathcal{Q})$, and
	$D_B = -\log \Gamma(\mathcal{P},\mathcal{Q})$.
\end{enumerate}

Given all the divergence measures in Equations (\ref{eq:div_l1}) -- (\ref{eq:div_js}), the diversity of any discrete distribution from
$\{\mathcal{X}_i\}_{i=1}^n$ can be measured as follows:
\begin{enumerate}
\item Replace $\mathcal{P}$ in Equations (\ref{eq:div_l1}) -- (\ref{eq:div_js}) with $\mathcal{X}_i$.
\item Replace $\mathcal{Q}$ in Equations (\ref{eq:div_l1}) -- (\ref{eq:div_js}) with the reference distribution, whether it be 
$\mathcal{M}(\bogs{\widehat{\pi}}^*)$, or $\mathcal{U}$ from Equation (\ref{eq:uniform}) if $\mathcal{M}(\bogs{\widehat{\pi}}^*)$ is not available.
\end{enumerate}
Since these particular divergences are analogous to distance measures, the smaller the divergence, the more diverse is the discrete 
distribution $\mathcal{X}_i$ with respect to the reference distribution of choice.

\subsection{Properties of Divergence--based Diversity Measures}
\label{subsec:prop_divergence_diversity}

Consider now how the proposed approach for measuring diversity differs from entropy measures with regards to the three problems 
introduced in \S~\ref{sec:introduction} for comparing the diversity of multiple communities.

First, using the set $\Omega^*$, we have a fixed unified set of species (or OTUs) for comparing all the samples.
This eliminates one source of variation among all the samples, and renders the difference between samples to be based only on the difference 
between their distributions.

Second, it can be noticed from all the divergence measures in Equations (\ref{eq:div_l1}) -- (\ref{eq:div_js}) that, zero elements in any 
distribution $\mathcal{X}_i$ penalizes the divergence between $\mathcal{X}_i$ and the reference distribution (whether it be 
$\mathcal{M}(\bogs{\widehat{\pi}}^*)$ or $\mathcal{U}$), and hence increases the divergence.
This is unlike entropy measures which ignores these zero elements.

Third, except for $D_\mathrm{SKL}$ and $D_\mathrm{JS}$, all other divergence measures do not impose any weighting scheme on the 
distribution $\mathcal{X}_i$.
For $D_\mathrm{SKL}$ in Equation (\ref{eq:div_skl}), the imposed weights $(p_j - q_j)$, are the differences between the probabilities for 
each outcome, which is maximized when the distributions are in complete disagreement, and zero when the distributions match.
This weighting scheme penalizes the difference (or disagreement) between the two distributions.
For $D_\mathrm{JS}$ in Equation (\ref{eq:div_js}), both distributions $\mathcal{P}$ and $\mathcal{Q}$ are compared against the middle 
distribution $\mathcal{Z}$.
If $\mathcal{P}$ completely disagrees with $\mathcal{Z}$, the difference $\log (p_j/z_j) = \log p_j - \log z_j$ is maximum, and it 
penalizes the final divergence $D_\mathrm{JS}$.
A similar interpretation follows for $\mathcal{Q}$ and $\mathcal{Z}$.
Here, it is important to note the difference between the weighting scheme for $\widehat{H}$ in Equation (\ref{eq:shannon_entropy}) on one 
hand, and that for $D_\mathrm{SKL}$ and $D_\mathrm{JS}$ on the other.
In $\widehat{H}$, the weights are set to create a balance between rare and common species, and hence they alter the original distribution 
of the sample.
However, in $D_\mathrm{SKL}$ and $D_\mathrm{JS}$ the weights penalize the disagreement (or the difference) between $\mathcal{X}_i$ and the 
reference distribution without altering any of them.

Divergence measures in general can be seen as distances between probability distributions.
However, unlike distance metrics which have measurement units, in information theory, divergence measures do not have such units.
Nevertheless, one cannot compare two different divergence values measured using two different divergence measures.
At this point, one may ask whether there is a biological interpretation for the divergence measures presented here.
Currently, from a statistical and information theoretic perspective, we cannot claim whether such an interpretation exist or not. 
If such an interpretation exists, it can be established by domain experts from each field through extensive analysis of these measures on 
their communities of interest.

Throughout the previous discussion we have always considered two reference distributions: 
(\emph{i}) the latent species distribution $\mathcal{M}(\bogs{\pi}^*)$, and
(\emph{ii}) the discrete uniform distribution $\mathcal{U}$.
In principle, we believe that any community $\mathcal{C}$ has its own latent species distribution $\mathcal{M}(\bogs{\pi}^*)$.
If an estimate for this distribution is available, say $\mathcal{M}(\bogs{\widehat{\pi}}^*)$, then one can use it as the reference 
distribution to measure the diversity of a community.
Due to their definition, entropy--based measures do not enjoy such a flexibility. 
When $\mathcal{M}(\bogs{\pi}^*)$ is not known, and hence $\mathcal{M}(\bogs{\widehat{\pi}}^*)$ is not available, one has no other option
but to use $\mathcal{U}$ as the reference distribution.
Still, divergence-based measures will be better to use for the three reasons mentioned above.
Another advantage of divergence-based measures is that they allow direct pairwise comparisons between all communities, which is not possible
to compute using entropy--based measures.

{\bf Invariance of ranking among groups.} 
When comparing the diversity of two communities, the ranking of the two communities should not be changed when a third community is added
to the comparison.
This is known as the invariance of ranking among groups. 
This property holds as well for divergence-based diversity measures under the condition that all groups have the same reference distribution.
If the reference distributions changes for one community, or for all communities, then the ranking among communities can change.
Note that this is a natural consequence of changing the reference distribution for one or all communities, and hence it should not be 
considered a disadvantage of divergence-based diversity measures.
Also note that it is not possible to compare the diversity of two or more communities with different reference distributions. 

{\bf Monotonicity and principle of transfer.}
For a community $\mathcal{C}$ with multinomial distribution $\mathcal{M}(m,\bogs{\pi})$, \citet{patil_taillie_1982} define the diversity of 
$\mathcal{C}$ as the average rarity $\delta(\mathcal{C}) = \sum_{i=1}^m \pi_i R(\pi_i)$, where $R(\pi_i)$ is the rarity of species $i$.
For instance, for Shannon's index $R(\pi_i) = -\log(\pi_i)$, while for Simpson's index $R(\pi_i) = (1 - \pi_i)$.
The rarity coefficient $R$ should satisfy two requirements:
\emph{(i)} $R$ is a nonnegative monotonic function, and
\emph{(ii)} $R$ satisfies the principle of transfer; i.e. diversity increases if a new species is introduced to the community, and/or
by making the distribution more even. 

Monotonicity is satisfied by the definition of divergence according to \citet{ali_silvey_66}, albeit in a different sense that suits the
nature of probability distributions.
Too see this, let $\div$ denote any of the previously mentioned divergence measures.
Then, by definition of divergence \citep{ali_silvey_66}, $\div(\mathcal{P},\mathcal{Q})$ is minimum when $\mathcal{P}=\mathcal{Q}$, 
and maximum when $\mathcal{P}$ and $\mathcal{Q}$ are orthogonal.
Further, let $\theta$ be a real parameter, and $\{\mathcal{P}_\theta ~ s.t. ~ \theta \in (a,b)\}$ be a family of mutually continuous 
distributions on the real line, such that $\mathcal{P}_\theta$ has a monotone likelihood ratio\footnote{Any two probability distributions
$\mathcal{P}(x)$ and $\mathcal{Q}(x)$ have the monotone likelihood ratio property if for any $x_1 > x_2$, we have that
$\mathcal{P}(x_1)/\mathcal{Q}(x_1) \geq \mathcal{P}(x_2)/\mathcal{Q}(x_2)$.
}. 
Then, if $a < \theta_1 < \theta_2 < \theta_3 < b$, we have that
$\div(\mathcal{P}_{\theta_1},\mathcal{P}_{\theta_2}) \leq \div(\mathcal{P}_{\theta_1},\mathcal{P}_{\theta_3})$. 
This property says that as the distance between the parameters (defining the distributions) increases, the divergence will increase as well.
This property immediately applies to our multinomial distributions parameterized with $(m,\bogs{\pi})$.

The principle of transfer, as explained above, has two aspects.
The first is that adding a new species to the community should increase the diversity. 
This property holds for all the proposed divergence since they are sums of individual coefficients, each representing one species.
The second is that increasing evenness should increase the diversity.
This property also holds for the proposed divergence measures when the reference distribution is the uniform distribution.
Increasing the evenness of a distribution makes it more similar to the uniform distribution, and hence decreases the divergence;
i.e. increases diversity.



\section{Concluding Remarks}
\label{sec:conclusion}

We propose a general methodology for measuring communities' diversity based on divergence measures.
Our work perceives diversity indices as measures for quantifying the difference (or discrepancy) between two probability distributions.
Entropy--based indices measure this difference in terms of similarity between a given distribution and the uniform distribution,
while divergence--based indices measure the difference between any two given distributions in a similar fashion to distances between points.
Our methodology retains all the properties of diversity indices, 
is flexible in terms of the reference distribution which all other communities will be compared with,
yields meaningful and comparable diversity values for samples withdrawn from different communities,
does not impose any weights on the species distribution, and 
allows for pairwise comparisons between all distributions. 
Further, it is easy to implement and is applicable to any community of interest.


\bibliographystyle{apalike}
\bibliography{sim_karim}
	


\end{document}